\documentclass[conference,letterpaper,10pt,twocolumn]{IEEEtran}
\usepackage[letterpaper, left=0.63in, right=0.63in, bottom=1in, top=0.7in]{geometry}
\usepackage[acronym]{glossaries}
\glsenableentrycount
\newacronym{WMMSE}{WMMSE}{weighted minimum mean square error}
\newacronym{FCL}{FCL}{fully connected layer}
\newacronym{NLOS}{NLOS}{non-line-of-sight}
\newacronym{LOS}{LOS}{line-of-sight}
\newacronym{MAC}{MAC}{multiply-accumulate}
\newacronym{BS}{BS}{base station}
\newacronym{PS}{PS}{phase-shifter}
\newacronym{NAS}{NAS}{neural architecture search}
\newacronym{PTQ}{PTQ}{post-training quantization}
\newacronym{QAT}{QAT}{quantization-aware training}
\newacronym{LSQ}{LSQ}{Learned Step-Size Quantization}
\newacronym{STE}{STE}{Straight Through Estimator}
\newacronym{MPQ}{MPQ}{mixed-precision quantization}
\newacronym{RL}{RL}{reinforcement learning}
\newacronym{AP}{AP}{analog precoder}
\newacronym{FC-HBF}{FC-HBF}{fully-connected HBF}
\newacronym{FSA-HBF}{FSA-HBF}{fixed subarray HBF}
\newacronym{DSA-HBF}{DSA-HBF}{dynamic subarray HBF}
\newacronym{BF}{BF}{beamforming}
\newacronym{UE}{UE}{user equipment}
\newacronym{AWGN}{AWGN}{additive white gaussian noise}
\newacronym{MIMO}{MIMO}{multiple-input multiple-output}
\newacronym{MISO}{MISO}{multiple-input single-output}
\newacronym{RF}{RF}{radio frequency}
\newacronym{RIS}{RIS}{reconfigurable intelligent surfaces}
\newacronym{IOT}{IOT}{internet-of-things}
\newacronym{QNN}{QNN}{quantized neural network}
\newacronym{CL}{CL}{convolutional layer}
\newacronym{UdeM-NLOS}{UdeM-NLOS}{Universite De Montreal-Non-Line-Of-Sight}
\newacronym{Okapark-LOS}{Okapark-LOS}{Okapark-Line-Of-Sight}
\newacronym{FDD}{FDD}{frequency division duplex}
\newacronym{TDD}{TDD}{time division duplex}
\newacronym{CSI}{CSI}{channel state information}
\newacronym{DNN}{DNN}{deep neural network}
\newacronym{DP}{DP}{digital precoder}
\newacronym{DL}{DL}{deep learning}
\newacronym{SVD}{SVD}{singular-value decomposition}
\newacronym{CNN}{CNN}{convolution neural network}
\newacronym{FDP}{FDP}{fully digital precoder}
\newacronym{SE}{SE}{spectral efficiency}
\newacronym{OFDM}{OFDM}{orthogonal frequency division multiplexing}
\newacronym{OMP}{OMP}{orthogonal matching pursuit}
\newacronym{FL}{FL}{fully-connected layer}
\newacronym{HSHO}{HSHO}{Hybrid Structured Heuristic Optimization}
\newacronym{HBF}{HBF}{hybrid beamforming}
\newacronym{IA}{IA}{initial access}
\newacronym{mm-Wave}{mm-Wave}{millimeter wave}
\newacronym{mMIMO}{mMIMO}{massive multiple-input multiple-output}
\newacronym{SINR}{SINR}{signal-to-interference-and-noise ratio}
\newacronym{SNR}{SNR}{signal-to-noise ratio}
\newacronym{EE}{EE}{Energy Efficiency}
\newacronym{RSSI}{RSSI}{received signal strength indicator}
\newacronym{PZF}{PZF}{phase zero forcing}
\newacronym{PSO}{PSO}{particle swarm optimization}
\newacronym{ZF}{ZF}{zero forcing}
\newacronym{O-FDP}{O-FDP}{optimal fully digital precoder}
\newacronym{JT}{JT}{joint transmission}
\newacronym{CU}{CU}{central unit}
\newacronym{MSE}{MSE}{mean square error}
\newacronym{CEL}{CEL}{cross entropy loss}
\newacronym{CB}{CB}{conjugate beamforming}
\newacronym{NC}{NC}{network controller}
\newacronym{CoMP}{CoMP}{coordinated multi point}
\newacronym{CF-mMIMO}{CF-mMIMO}{cell-free massive MIMO}
\newacronym{CF-HBF}{CF-HBF}{cell-free hybrid beamforming}
\newacronym{CF-BF}{CF-BF}{cell-free beamforming}
\newacronym{MLDG}{MLDG}{Meta-Learning Domain Generalization}
\newacronym{MAML}{MAML}{Model Agnostic Meta-Learning}
\newacronym{WSR}{WSR}{Weighted Sum Rate}
\newacronym{SwiGLU}{SwiGLU}{Swish-Gated Linear Unit}
\newacronym{MHSA}{MHSA}{Multi-Head Self-Attention}
\newacronym{FFN}{FFN}{Feed-Forward Network}
\newacronym{FPGA}{FPGA}{Field-Programmable Gate Array}
\newacronym{upa}{UPA}{uniform planar array}
\newacronym{ART}{ART}{automatic resolution tuning}
\newif\ifDeepMIMOModel
\DeepMIMOModeltrue

\newif\ifSimpleNParamEq
\SimpleNParamEqtrue

\usepackage{lettrine}
\usepackage[ruled,vlined,linesnumbered]{algorithm2e}
\usepackage{multirow}
\usepackage{longtable}
\usepackage[table,xcdraw]{xcolor}
\usepackage{array} 
\usepackage{subfigure}
\usepackage{stfloats}
\makeatletter
\let\oldlt\longtable
\let\endoldlt\endlongtable
\def\longtable{\@ifnextchar[\longtable@i \longtable@ii}
\def\longtable@i[#1]{\begin{figure}[t]
\onecolumn
\begin{minipage}{0.5\textwidth}
\oldlt[#1]
}
\def\longtable@ii{\begin{figure}[t]
\onecolumn
\begin{minipage}{0.5\textwidth}
\oldlt
}
\def\endlongtable{\endoldlt
\end{minipage}
\twocolumn
\end{figure}}
\makeatother
\usepackage{ifpdf}
\ifCLASSINFOpdf
  \usepackage[pdftex]{graphicx}
  \usepackage{graphicx,epstopdf}
  \graphicspath{{../pdf/}{../jpeg/}}
  \DeclareGraphicsExtensions{.pdf,.jpeg,.png,.eps}
\else
  \usepackage[dvips]{graphicx}
  \usepackage{graphicx,epstopdf}
  \graphicspath{{../eps/}}
  \DeclareGraphicsExtensions{}
\fi
\usepackage{tikz}
\usetikzlibrary{decorations.pathreplacing,calc}

\usepackage{cite}
\usepackage{amsthm}
\usepackage{steinmetz}
\usepackage{amssymb}
\usepackage{balance}
\usepackage{eqparbox}
\usepackage{multirow}
\usepackage{bbm}
\usepackage{float}
\SetAlFnt{\small}

\usepackage{amsmath}

\usepackage{mathtools, nccmath}

\usepackage{booktabs, siunitx}
\usepackage{dcolumn}
\newcolumntype{d}[1]{D{.}{.}{#1}}
\usepackage[scaled]{DejaVuSansMono}
\usepackage[T1]{fontenc}

\usepackage{color}
\usepackage{relsize}
\usepackage{mathtools}

\usepackage{mathtools}

\SetKwInput{KwInput}{Input}                
\SetKwInput{KwOutput}{Output}
\SetKwInput{KwOutputr}{Output Regression}              
\SetKwInput{KwOutputc}{Output Classification}              

\newcommand{\bseq}{\begin{subequations}}
\newcommand{\eseq}{\end{subequations}}
\newcommand{\baln}{\begin{align}}
\newcommand{\ealn}{\end{align}}
\newcommand{\balnd}{\begin{aligned}}
\newcommand{\ealnd}{\end{aligned}}
\newcommand{\beq}{\begin{equation}}
\newcommand{\eeq}{\end{equation}}
\newcommand{\beqn}{\begin{eqnarray}}
\newcommand{\eeqn}{\end{eqnarray}}
\newcommand{\beqno}{\begin{eqnarray*}}
\newcommand{\eeqno}{\end{eqnarray*}}
\newcommand{\bma}{\begin{displaymath}}
\newcommand{\ema}{\end{displaymath}}
\newcommand{\bnu}{\begin{enumerate}}
\newcommand{\enu}{\end{enumerate}}
\newcommand{\bce}{\begin{center}}
\newcommand{\ece}{\end{center}}
\newcommand{\btb}{\begin{tabular}}
\newcommand{\etb}{\end{tabular}}
\newcommand{\ba}{\begin{array}}
\newcommand{\ea}{\end{array}}
\usepackage{footnote}
\makesavenoteenv{tabular}
\makesavenoteenv{table}
\makeatletter 
\newcommand\semiHuge{\@setfontsize\semiHuge{21.1}{27.38}}
\makeatother

\usepackage{fix-cm}
\usepackage{cite}
\usepackage{amsmath,amssymb,amsfonts}
\usepackage{algorithmic}
\usepackage{color}
\usepackage{textcomp}
\usepackage{float}
\usepackage[T1]{fontenc}
\usepackage{mathptmx} 
\begin{document}

\title{Quantized Transformers for Massive MIMO Precoding with Automatic Resolution Tuning}

\author{\IEEEauthorblockN{Ghazal~Kasalaee,
        Glodi Sala Mangituka,
        Ali~Hasanzadeh~Karkan,
        \\Jean-François~Frigon,
        and~François~Leduc-Primeau}
\IEEEauthorblockA{Department of Electrical Engineering, Polytechnique Montréal, Montréal, QC H3C 3A7, Canada\\
Emails: \{ghazal.kasalaee, glodi.sala-mangituka, ali.hasanzadeh-karkan, j-f.frigon, francois.leduc-primeau\}@polymtl.ca
    }
}

\maketitle

\begin{abstract}
Deep learning precoders, particularly those based on Transformer architectures, offer superior spectral efficiency for Massive MIMO systems but may incur high computational cost. While existing mixed-precision quantization studies on convolutional neural networks establish important energy baselines, traditional discrete search methods fail to scale to the large parameter spaces of modern Transformers. To bridge this gap, we apply a differentiable precision learning technique to massive MIMO precoding. This approach autonomously and jointly optimizes weights, activations, quantization step sizes, and layer-wise bit-widths in a single training loop, directly optimizing for energy efficiency. We further expand this approach by incorporating Neural Architecture Search across various Transformer model sizes and evaluating the impact of random versus floating-point trained initializations on resolution tuning. Ultimately, we demonstrate that the proposed training method enables the deployment of highly compact Transformer precoders, improving energy efficiency by up to 288$\times$ compared to the classical Weighted Minimum Mean Square Error algorithm at equal sum-rate performance in a dense downtown environment. 
\end{abstract}



\section{Introduction} \label{Sec:Intro}
As mobile networks evolve toward an AI-native 6G air interface, the radio access network faces a dual challenge: maximizing spectral efficiency while minimizing its energy footprint, which currently accounts for 70\% of network power consumption~\cite{gsma2021}. \Gls{mMIMO} baseband processing is a primary contributor to this load. While classical algorithms such as \gls{WMMSE}~\cite{shi2011iteratively} are robust, their computational complexity scales cubically with the number of transmit antennas, making them increasingly unsustainable for massive arrays. Consequently, \glspl{DNN} have emerged as a compelling alternative. Compared to traditional methods, they can achieve superior trade-offs between compute energy and system performance by learning efficient approximations.

To address the escalating complexity and power consumption of traditional precoding, the literature has increasingly focused on energy-efficient \glspl{DNN} tailored to hardware constraints. One strategy involves designing highly efficient precoder structures, such as using unsupervised \glspl{DNN} to discover near-optimal, energy-efficient precoders~\cite{karkan2025low}. Similarly, dedicated convolutional frameworks have been proposed to optimize phase-shift matrices under hardware constraints, maximizing energy efficiency~\cite{liu2022dl}. To further reduce implementation complexity on resource-constrained devices, substantial research has shifted toward low-resolution hardware and quantized \glspl{DNN}. Works such as~\cite{he2020model} have developed DL-based precoders specifically for systems employing low-resolution digital-to-analog converters. The approach in~\cite{bo2020deep} uses \glspl{CNN} to predict analog precoders tailored for low-resolution phase shifters, while~\cite{ma2020sparse} introduces neural networks trained offline that directly account for approximate phase quantization. Pushing quantization further into the neural network architecture itself, \cite{kumar2023multi} relies on fixed 8-bit quantization and offline target generation to reduce the hardware complexity of the \gls{CNN}-based hybrid precoder.

Recent advancements in physical layer processing have seen a shift toward Transformer-based architectures~\cite{Emery2025}, which unlock new levels of precoding accuracy by capturing long-range channel dependencies. However, these performance gains introduce higher computational demands compared to earlier convolutional models, potentially offsetting the energy benefits of AI-native processing. While robust efficiency baselines for \gls{CNN}-based precoders have been established using \cgls{MPQ} guided by \cgls{NAS}~\cite{Kasalaee2025}, extending traditional search methods to Transformers faces a scalability hurdle. Specifically, the combinatorial explosion of bit configurations across multi-head attention layers makes exhaustive search computationally prohibitive, creating a critical need for new methods that bridge the gap between high-accuracy models and practical deployment on energy-constrained \glspl{BS}.

In this work, we aim to maximize the compute efficiency of wireless-communication \glspl{DNN} by targeting \gls{mMIMO} deployment on precision-scalable hardware. While Transformers are typically computationally demanding, those designed for physical layer processing remain sufficiently compact to allow \glspl{FPGA} to instantiate dedicated blocks tailored to the precision of each layer. Exploiting this flexibility requires optimal bit-width configurations. However, the massive parameter spaces of modern Transformers make traditional discrete search methods prohibitive. To address this, we present the first application of an \cgls{ART} framework to wireless precoding. By relaxing bit-width selection into a continuous variable, the proposed training method jointly optimizes network weights, activations, quantization step sizes, and layer-wise bit-width assignments in a single training run. Leveraging \gls{LSQ}~\cite{Esser2020} and embedding a hardware-calibrated energy model directly into the loss function, the framework autonomously learns highly efficient precision maps. We further consider several Transformer model sizes and systematically evaluate the best combination of floating-point pre-training and \cgls{QAT} on model performance.

\section{System Model} \label{Sec:sysmodel}

We consider a downlink \gls{mMIMO} system operating in \cgls{FDD} mode, where a \gls{BS} equipped with $N_{\sf T}$ antennas serves $N_{\sf U}$ single-antenna users. Let $\mathbf{h}_u \in \mathbb{C}^{N_{\sf T} \times 1}$ denote the channel vector for user $u$. The received signal $y_u$ is given by
\begin{equation}
    y_u = \mathbf{h}_u^H \sum_{k=1}^{N_{\sf U}} \mathbf{w}_{p,k} x_k + n_u,
\end{equation}
where $\mathbf{w}_{p,k} \in \mathbb{C}^{N_{\sf T} \times 1}$ is the precoding vector for user $k$, $x_k$ is the data symbol, and $n_u \sim \mathcal{CN}(0, \sigma^2)$ is additive white Gaussian noise.
The corresponding \gls{SINR} for user $u$ is formulated as
\begin{equation} \label{eq:sinr_system}
    \text{SINR}_u = \frac{|\mathbf{h}_u^H \mathbf{w}_{p,u}|^2}{\sum_{j \neq u} |\mathbf{h}_u^H \mathbf{w}_{p,j}|^2 + \sigma^2}.
\end{equation}

The system aims to maximize the downlink sum-rate $R(\mathbf{W}_p) = \sum_{u} \log_2(1 + \text{SINR}_u)$ under a total power constraint $\sum_u \|\mathbf{w}_{p,u}\|^2 \leq P_{\max}$.

\subsection{Dataset Generation and Channel Modeling} \label{Dataset}
To accurately capture the spatial channel characteristics inherent to practical precoding in \gls{mMIMO} systems, we generated custom, site-specific datasets using the \textsc{matlab} ray-tracing toolbox. Leveraging environmental data from OpenStreetMap~\cite{OpenStreetMap} to ensure physical realism, the simulated \glspl{BS} were virtually deployed in two distinct scenarios within the Montreal region. The ``Ericsson'' site captures an urban industrial environment where 75\% of users experience \gls{LOS} conditions. In contrast, the ``Sainte-Catherine'' site represents a dense downtown environment featuring shorter link distances, stronger blockage variability, and 75\% \gls{NLOS} links.

For both scenarios, the \gls{BS} is equipped with an $8 \times 8$ \cgls{upa} featuring half-wavelength antenna spacing and operating at a carrier frequency of $2$\,GHz. The transmitter is placed at a height of $20$\,m with an output power of $20$\,W, accounting for a $10$\,dB system loss. User equipment was distributed in concentric circular patterns around the \gls{BS} at radial distances ranging from $50$\,m to $350$\,m, separated by $10^\circ$ angular intervals. This spatial distribution provides a robust representation of varied deployment scenarios and localized channel fading conditions. To model urban multipath propagation, the ray-tracing engine computed up to $10$ reflections off surrounding buildings and terrain.

\section{Methodology} \label{Sec:Proposed}
Unlike~\cite{Kasalaee2025}, which used discrete \gls{NAS} to find optimal bit configurations, this work leverages \cgls{ART}, a fully differentiable framework that learns precision levels via gradient descent.

\subsection{Transformer Neural Architecture}\label{sec:DNNarch}
Our underlying model is a Stacked Transformer tailored for wireless physical-layer processing. The network accepts the concatenated real and imaginary components of the \gls{CSI} matrix as input and predicts the complex beamforming precoding matrix. The architecture consists of an initial embedding layer, followed by a series of Transformer blocks, and a final linear projection layer. Each Transformer block utilizes \cgls{MHSA} and a \gls{FFN} equipped with \gls{SwiGLU} activation functions. Residual connections are employed around both of these sub-layers to facilitate gradient flow. Furthermore, Layer Normalization is applied immediately following each residual addition.

To explore the trade-off between compute energy and system performance, we define five distinct model capacities ranging from ``Tiny'' to ``Large.'' These variants are generated by scaling four architectural hyperparameters: the embedding dimension $D$ ($64$ to $256$), the number of attention heads $H$ ($2$ to $8$), the FFN hidden dimension $D_{ff}$ ($512$ to $4096$), and the number of Transformer layers $L$ ($2$ to $8$). This scaling yields models ranging from approximately 0.5M to 25M parameters, allowing us to map the Pareto front across diverse hardware constraints.

\subsection{Hardware-Aware Energy Modeling}
To take into account the inference computation cost during training, we formulate a hardware-aware energy model adapted to the considered Transformer architectures. Building upon \cite{moons:2017a}, we decompose the total hardware inference energy of a layer $l$ into computation ($E_{\text{C}}^l$), weight transfer ($E_{\text{W}}^l$), and activation transfer ($E_{\text{A}}^l$). 

As will be discussed further in Section~\ref{sec:ART}, we consider that weights in layer $l$ are represented using $\tilde{b}^l_w$ bits, and activations using $\tilde{b}^l_a$ bits.
Calibrated against 45-nm CMOS empirical measurements \cite{Horowitz2014}, the compute energy is modeled as $E_{\text{MAC}}(\tilde{b}^l) = \alpha \cdot (\tilde{b}^l/16)^\beta$, with constants $\alpha = 0.8579$ and $\beta = 1.9$. 
To simplify the model, we define the memory access energy in terms of  $E_{\text{MAC}}$.
A main memory access incurs 
$E_{\text{M}}(\tilde{b}) = 2\,E_{\text{MAC}}(\tilde{b})$, while local 
register access costs $E_{\text{L}}(\tilde{b}) = E_{\text{MAC}}(\tilde{b})$, 
substituting $\tilde{b}^l_w$ or $\tilde{b}^l_a$ depending on whether 
weights or activations are being transferred.

\begin{table}[t]
\centering
\caption{Architectural Complexity}
\label{tab:complexity}
\resizebox{\columnwidth}{!}{
\begin{tabular}{@{}lccc@{}}
\toprule
\textbf{Component} & \textbf{MAC Operations ($N_{\text{MAC}}$)} & \textbf{Weights ($N_{\text{W}}$)} & \textbf{Activations ($N_{\text{A}}$)} \\ \midrule
Embedding & $2N_{\sf T} N_{\sf U} D$  & $2N_{\sf T} D$ & $N_{\sf U} D$ \\ \midrule
\multicolumn{4}{@{}l}{\textbf{Per Transformer Block ($\times L$):}} \\ \addlinespace[2pt]
\quad \gls{MHSA} & $4N_{\sf U}D^2 + 2N_{\sf U}^2D$ & $4D^2$ & $4N_{\sf U}D$ \\
\quad \gls{SwiGLU} \gls{FFN} & $3N_{\sf U} D D_{\text{ff}}$ & $3D D_{\text{ff}}$ & $N_{\sf U} D_{\text{ff}}$ \\
\midrule

Output Projection      & $2N_{\sf T} N_{\sf U} D$  & $2N_{\sf T} D$ & $2N_{\sf U} N_{\sf T}$ \\ \bottomrule
\end{tabular}}
\end{table}

The inference complexity of the Transformer architecture, parameterized by the system parameters introduced in Section~\ref{Sec:sysmodel} and the model-architecture parameters introduced in Section~\ref{sec:DNNarch}, is summarized in Table~\ref{tab:complexity}. The \gls{MHSA} footprint stems from four $D \times D$ linear projections (Query, Key, Value, and Output), contributing $4N_{\sf U}D^2$ MACs, alongside the core self-attention operations, which add $2N_{\sf U}^2D$ MACs. Similarly, the \gls{SwiGLU} \gls{FFN} uses three distinct $D \times D_{\text{ff}}$ projection matrices for its gating mechanism. Using these derived hardware counts, the total computational energy accounts for both the linear operations and an element-wise overhead ($3N_{\text{A}}$) for layer normalization, bias, and activation:
\begin{equation}
    E_{\text{C}}^l(\tilde{b}^l_w, \tilde{b}^l_a) = E_{\text{MAC}}(\tilde{b}^l_w)\,N_{\text{MAC}} +3\,E_{\text{MAC}}(\tilde{b}^l_a)\,N_{\text{A}}.
\end{equation}

Still following \cite{moons:2017a}, the number of local buffer accesses is determined by taking into account data reuse opportunities arising from parallel processing, in this case assuming 64 processing elements. We thus define parallelism factors $p_w^l = 64 \times (16/\tilde{b}_w^l)$ and $p_a^l = 64 \times (16/\tilde{b}_a^l)$ and define the memory access energy for weights and activations as
\begin{align}
    E_{\text{W}}^l(\tilde{b}^l_w) &= E_{\text{M}}(\tilde{b}^l_w) \, N_{\text{W}} + E_{\text{L}}(\tilde{b}^l_w) \, \frac{N_{\text{MAC}}}{\sqrt{p_w^l}}, \\
    E_{\text{A}}^l(\tilde{b}^l_a) &= 2 \, E_{\text{M}}(\tilde{b}^l_a) \, N_{\text{A}} + E_{\text{L}}(\tilde{b}^l_a) \, \frac{N_{\text{MAC}}}{\sqrt{p_a^l}}.
\end{align}

Finally, the layer-wise components are aggregated across all $L$ layers to yield the total network inference energy. To maintain differentiability during training, the total energy is evaluated directly on the continuous precision vector $\tilde{\mathbf{b}}_w$ and $\tilde{\mathbf{b}}_a$:
\begin{equation} \label{eq:energy_model}
    E(\tilde{\mathbf{b}}_w,~\tilde{\mathbf{b}}_a) = \sum_{l=1}^{L} \Big( E_{\text{C}}^l(\tilde{b}^l_w,\tilde{b}^l_a) + E_{\text{W}}^l(\tilde{b}^l_w) + E_{\text{A}}^l(\tilde{b}^l_a) \Big).
\end{equation}
By rendering $E_{\text{W}}^l$ and $E_{\text{A}}^l$ as a smooth function of $\tilde{b}^l_w$ and $\tilde{b}^l_a$, respectively, gradient descent autonomously compresses the spatial memory footprint without requiring heuristic precision tuning.

\subsection{Automatic Resolution Tuning and Joint Optimization}\label{sec:ART}
The training objective is a composite loss function designed to maximize spectral efficiency while autonomously minimizing computational energy via gradient descent. 
Each layer is assigned a continuous precision parameter $\tilde{b}^l_w$ and $\tilde{b}^l_a$. Inspired by the continuous domain extension introduced in \cite{nikolic2024bitpruning}, the respective tensors are quantized via a differentiable interpolation between two adjacent integer precisions:
\begin{align} \label{eq:bit_interp}
    \widehat{X}^l = (1-\gamma^l)\,Q(X^l;\lfloor\tilde{b}^l\rfloor) + \gamma^l\,Q(X^l;\lfloor\tilde{b}^l\rfloor+1),
\end{align}
where $\gamma^l = \tilde{b}^l - \lfloor\tilde{b}^l\rfloor$ acts as the interpolation weight, and $Q(X^l; b)$ denotes \gls{LSQ} quantization at $b$ bits, applied separately to weights with $b=\tilde{b}^l_w$ and to activations 
with $b=\tilde{b}^l_a$. The floor operation utilizes a straight-through estimator during the backward pass to preserve gradient flow.

The achievable sum-rate becomes a function of these precision variables, defining the task loss:
\begin{equation} \label{eq:task_loss}
    \mathcal{L}_{\text{task}}(\tilde{\mathbf{b}}_w, \tilde{\mathbf{b}}_a) = 
    -\sum_{u=1}^{N_U} \log_2\!\big(1 + 
    \text{SINR}_u(\tilde{\mathbf{b}}_w, \tilde{\mathbf{b}}_a)\big).
\end{equation}
The total loss combines $\mathcal{L}_{\text{task}}$ with the hardware-calibrated energy penalty from \eqref{eq:energy_model}:
\begin{equation} \label{eq:total_loss}
    \mathcal{L}(\tilde{\mathbf{b}}_w, \tilde{\mathbf{b}}_a) = 
    \frac{\mathcal{L}_{\text{task}}(\tilde{\mathbf{b}}_w, \tilde{\mathbf{b}}_a)}
    {\overline{R}_{\mathrm{FP32}}} + \lambda\,
    \frac{E(\tilde{\mathbf{b}}_w,\tilde{\mathbf{b}}_a)}{E_{\mathrm{8bit}}},
\end{equation}
where $\overline{R}_{\mathrm{FP32}}$ is the sum-rate achieved by a full-precision model of the same size, and $E_{\mathrm{8bit}}$ is the model-size-specific inference energy under uniform 8-bit quantization. This normalization renders both terms dimensionless, allowing the trade-off parameter $\lambda$ to control the accuracy-energy balance uniformly across model sizes.

\subsection{Training Protocol}\label{sec:training_protocol}
Each model is trained independently on a single deployment site using the AdamW optimizer with a cosine annealing learning rate schedule. Training proceeds in three phases. Initially, the model is pretrained exclusively on the task loss, $\mathcal{L}_{\text{task}}$, for up to 300 epochs using a model-size-dependent learning rate. This initial phase establishes $\overline{R}_{\mathrm{FP32}}$ and $E_{\mathrm{8bit}}$ as normalization anchors. 

The second phase transitions to a joint optimization of network weights, quantizer step sizes, and continuous precision parameters, $\tilde{b}^l_w, \tilde{b}^l_a\in [1, 16]$, initialized to $4$. This joint training spans 120 epochs using the composite loss function defined in \eqref{eq:total_loss} with a fixed trade-off parameter $\lambda$. During this phase, the weights maintain the size-dependent learning rate from pretraining, while the bit parameters are updated via a fixed, model-size-independent learning rate. 

In the final phase, each $\tilde{b}^l_w$ and $\tilde{b}^l_a$  is rounded to its nearest integer and frozen. The weights and quantizer step sizes are then fine-tuned for an additional 30 epochs solely on $\mathcal{L}_{\text{task}}$ to close the performance gap between fractional training precision and discrete hardware deployment. To map the Pareto front, this entire end-to-end procedure is repeated as $\lambda$ is swept across the set $\{0.1, 0.5, 1, 5, 10, 25, 50\}$, yielding a uniquely optimized model for each constraint value.

\section{Numerical Results} \label{Sec:Simulation}
\begin{figure*}[ht!]
    \centering
    \subfigure[Ericsson site]{\includegraphics[width=0.95\columnwidth]{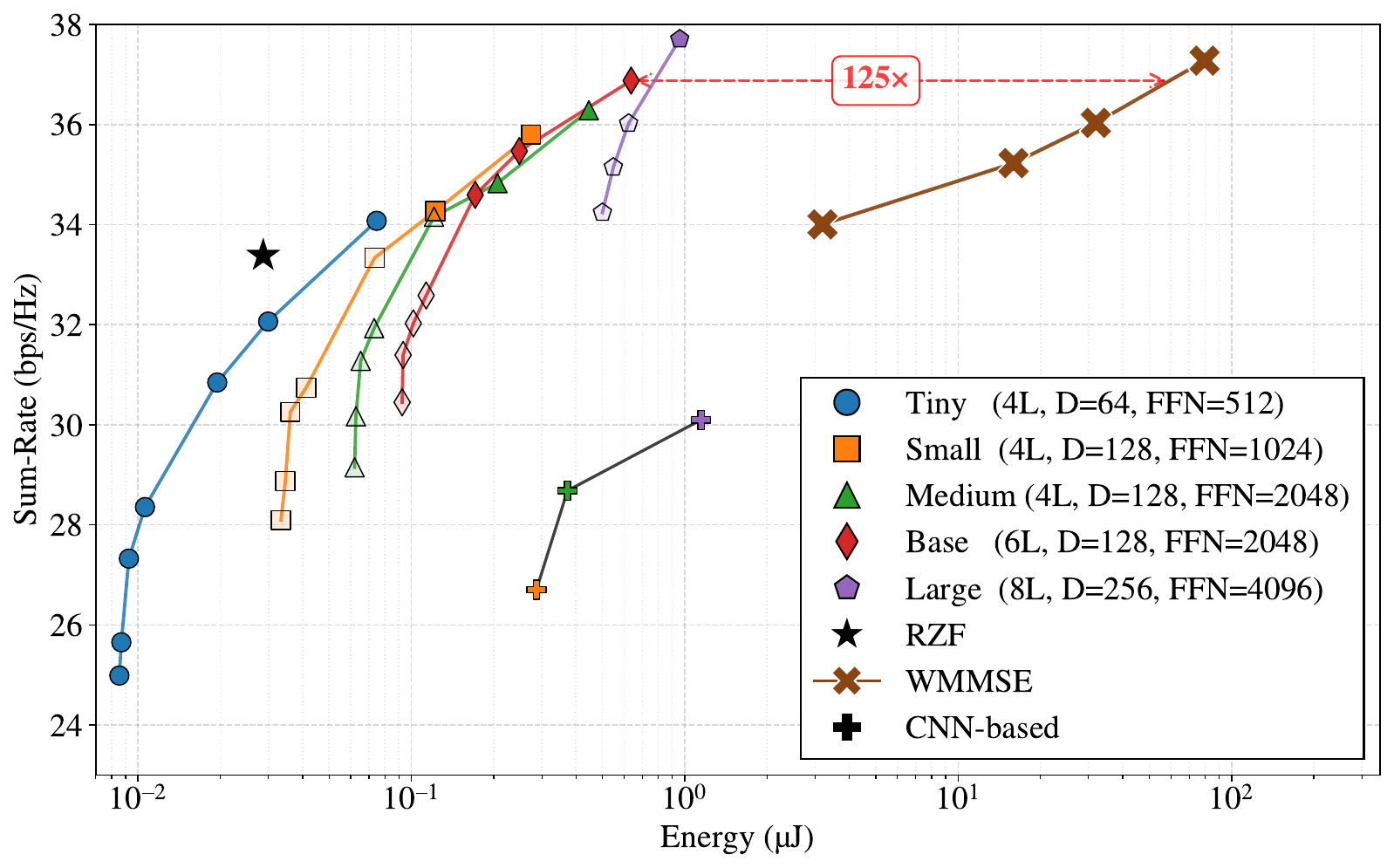}
    \label{fig:pareto-Ericsson}}
    \hfill
    \subfigure[Sainte-Catherine site]{\includegraphics[width=.95\columnwidth]{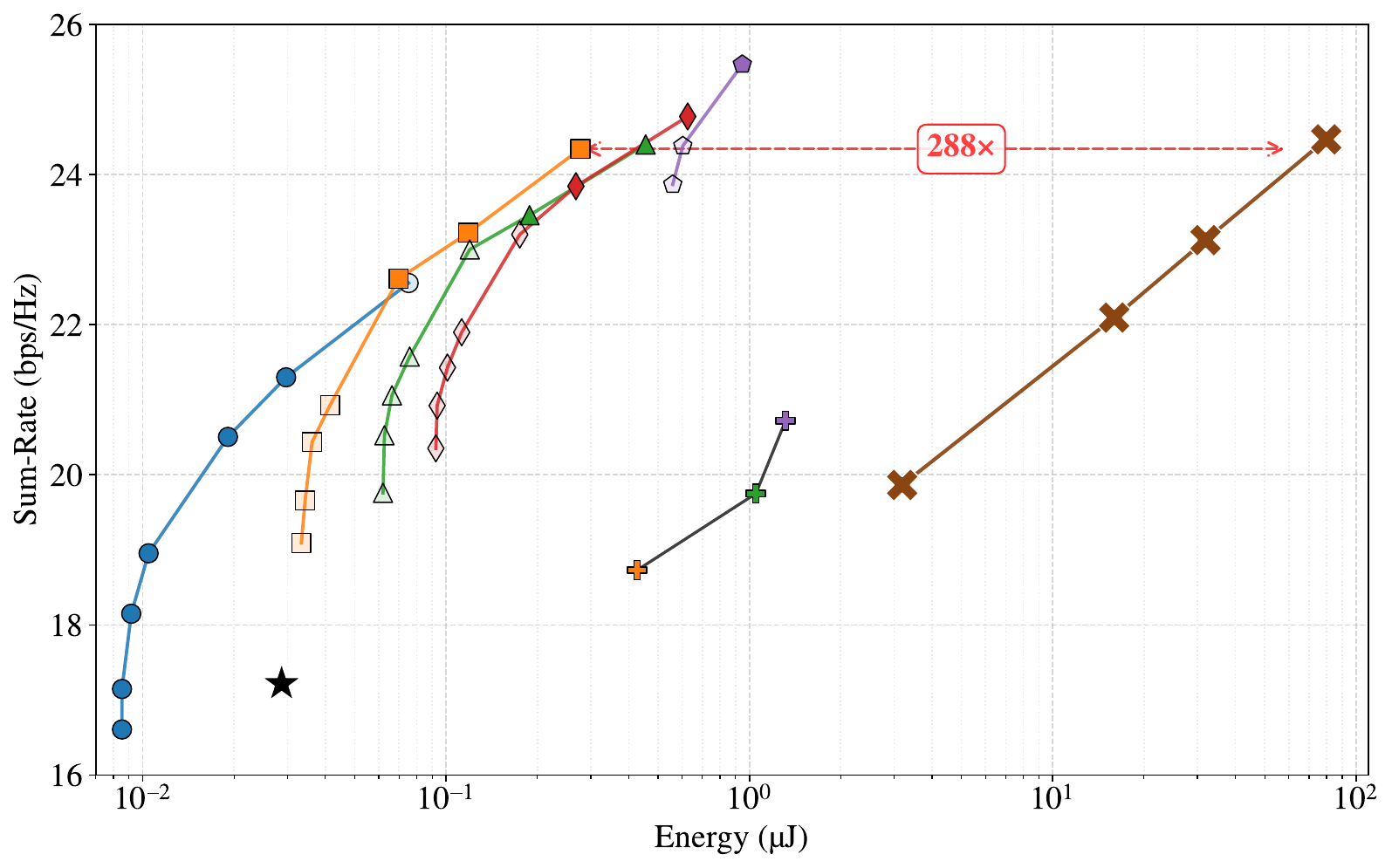}
    \label{fig:pareto-Saint}}
    \caption{Sum-rate versus inference energy for five Transformer models of varying sizes trained with ART on two deployment sites, with $\lambda \in \{0.1, 0.5, 1, 5, 10, 25, 50\}$. Filled markers form the Pareto front; hollow markers are dominated. ART training is also applied to a CNN-based precoder ($+$). RZF ($\bigstar$) and WMMSE ($\times$, $I \in \{1, 5, 10, 25\}$ iterations) are shown as baselines. Horizontal arrows indicate energy gains at equal sum-rate relative to WMMSE.}\label{fig:pareto}
\end{figure*}
\begin{figure*}[t!]
    \centering
    \subfigure[Ericsson site]{\includegraphics[width=\columnwidth]{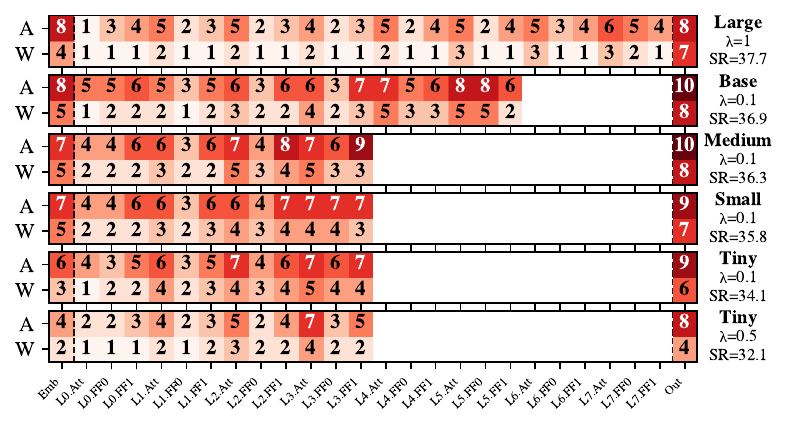}
    \label{fig:precision-W}}
    \subfigure[Sainte-Catherine site]{\includegraphics[width=\columnwidth]{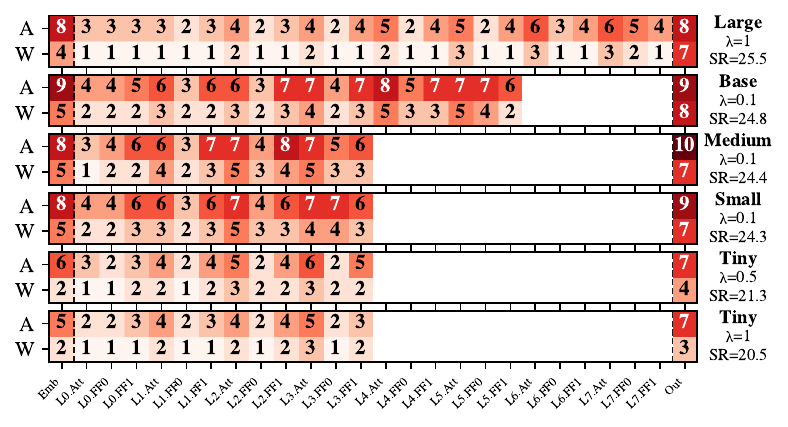}
    \label{fig:precision-A}}
    \caption{Layer-wise bit-widths (W: weights, A: activations) learned by ART for six model configurations $(\text{model},\,\lambda)$ appearing on the Pareto front in Fig.~\ref{fig:pareto}. Achieved sum-rates (bps/Hz) are reported in each row. As the rate--energy trade-off tightens, weight resolutions are progressively reduced to 1--2 bits in intermediate layers, while activations and output projections retain higher precision.}
    \label{fig:precision}
\end{figure*}
We train Transformer-based precoders across five architectural footprints from Tiny (4 layers, $D=64$, 0.2M parameters) to Large (8 layers, $D=256$, 18.6M parameters), each evaluated across $\lambda \in \{0.1, 0.5, 1, 5, 10, 25, 50\}$. We compare against two classical baselines: regularized zero-forcing (RZF)~\cite{sanguinetti2014optimal} and the \gls{WMMSE} algorithm evaluated at $I \in \{1, 5, 10, 25\}$ iterations, alongside a CNN-based precoder trained with ART. The energy for the classical baselines was evaluated using the same model as in~\cite{Kasalaee2025}.

The CNN precoder baseline is constructed using a residual convolutional pipeline comprising three convolutional layers for localized spatial feature extraction, each with $C_{\text{out}}$ output channels, followed by three \glspl{FCL} of size $D_{\text{FCL}}$ and two parallel linear readout heads that map the real and imaginary components of the precoding matrix. To trace the efficiency Pareto front, this baseline is evaluated across three capacities: Small (1.3M parameters, $C_{\text{out}}=16$, $D_{\text{FCL}}=256$), Medium (5.0M parameters, $C_{\text{out}}=32$, $D_{\text{FCL}}=512$), and Large (19.5M parameters, $C_{\text{out}}=64$, $D_{\text{FCL}}=1024$). To ensure a fair comparison, all CNN variants are trained in the same way as the Transformer models, with $\lambda=1$.

\subsection{Pareto Efficiency Analysis}
Fig.~\ref{fig:pareto} compares the downlink sum-rate and hardware inference energy. The proposed Transformer models can exceed the peak sum-rate of WMMSE ($I=25$) while consuming $125\times$ less compute energy at the Ericsson site and $288\times$ less at the Sainte-Catherine site. Across all evaluated capacity constraints, the Transformer configurations consistently dominate the CNN-based precoders, which confirms the advantage of the attention mechanism for spatial multiplexing.

On the Ericsson site, \gls{LOS} conditions make RZF a highly competitive reference for energy efficiency, slightly outperforming the Transformer models in energy consumption, albeit at a sum-rate about 10\% lower than the best-performing precoders. Note that this gap could potentially be reduced by adding another Transformer model size below ``Tiny''.

Conversely, on the Sainte-Catherine site, where rich scattering and blockage dominate, the aggressively compressed Tiny model matches RZF throughput at $3.4\times$ lower energy.

While our training framework proves highly effective at compressing individual architectures (e.g., increasing $\lambda$ from 0.1 to 5 for the Base model yields a $5.6\times$ energy reduction), expanding our analysis to the Pareto front reveals a broader architectural design rule: at a fixed energy budget, a smaller model with moderate compression consistently outperforms a larger, aggressively quantized alternative.

Thanks to the normalization introduced in the loss function in \eqref{eq:total_loss}, a single choice of the hyperparameter $\lambda$ is broadly applicable for generating models on the Pareto front. For the Transformer models, $\lambda=0.1$ almost always yields a model on the Pareto front.

In addition to their energy efficiency, Transformer-based precoders also provide practical advantages, such as supporting variable per-user rate requirements~\cite{Emery2025} or maintaining good performance in the presence of channel estimation errors~\cite{ali-arxiv}.

\begin{figure}[t!]
    \centering
    \includegraphics[width=\columnwidth]{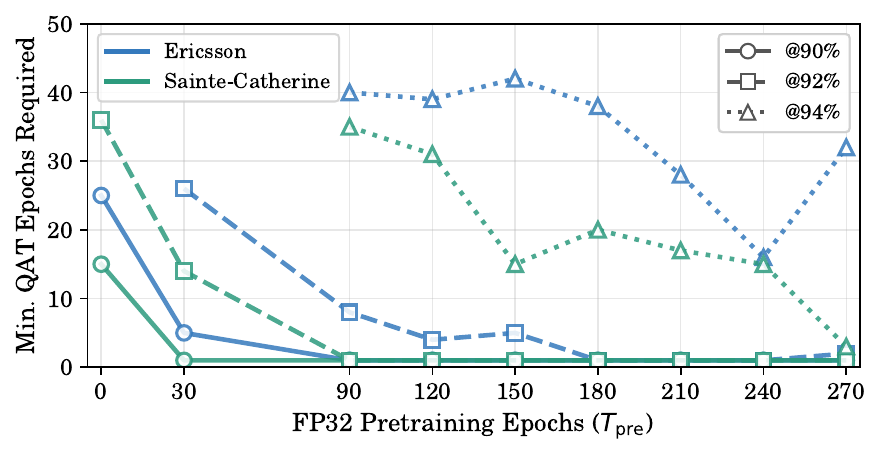}
    \vspace{-20pt}
    \caption{Impact of Phase~1 (FP32) pretraining budget on quantization robustness and the minimum QAT epochs required to achieve specific thresholds.}
    \vspace{-10pt}
    \label{fig:QATepochs}
\end{figure}

\subsection{Bit-width Allocations}
Fig.~\ref{fig:precision} provides a complete view of the learned layerwise bit-widths for some of the configurations on the Pareto front, across both deployment sites.
We first note that the optimization process autonomously finds that the first and last layers should retain higher precision, which is often used as a rule of thumb when training quantized models.
Furthermore, we observe that weights can be quantized more aggressively than activations. In fact, the weight resolution is often reduced to only 1 or 2 bits. Across all trained models, the largest resolutions assigned to weights and activations are 8 and 10 bits, respectively.
These results showcase the need for precision-scalable hardware architectures. On the one hand, they highlight the value of dedicated FPGA acceleration for these models, enabling arbitrary bit-width assignments tailored to a desired energy budget. On the other hand, they inform the choice of supported bit-widths in ASIC accelerators.

\subsection{FP32 Pretraining Budget}
Finally, we present an additional experiment to motivate the two-phase training protocol presented in Section~\ref{sec:training_protocol}.
Since we need to train multiple models of the same size for different values of $\lambda$, we can reduce the overall training budget by sharing a pretraining phase across all $\lambda$ values.
For example, considering the ``Small'' Transformer with $\lambda=5$, Fig.~\ref{fig:QATepochs} illustrates the minimum number of QAT epochs required to reach various performance thresholds (90\%, 92\%, and 94\% of $\overline{R}_\mathrm{FP32}$).
If the same model is instead trained from scratch using only QAT, about 300 epochs are required for convergence.
Therefore, FP pretraining significantly reduces the overall training budget when training more than one quantized model.

\section{Conclusion} \label{Sec:Conclusion}
In this paper, we investigated training Transformer-based \gls{mMIMO} precoding models using advanced quantization-aware methods. Our training recipe combined automatic tuning of the weight and activation bit-widths with learned quantization step sizes, while pursuing a dual objective of sum-rate maximization and compute-energy minimization.
This approach eliminates the combinatorial complexity of searching for the best bit-width assignment at each layer and instead automatically allocates precision only where necessary to maintain spectral efficiency.
Our evaluations on two distinct ray-tracing channel models demonstrate that the proposed training pipeline generates Pareto-efficient models with little hyperparameter tuning. Notably, we identify a mixed-precision model that achieves a $288\times$ improvement in energy efficiency compared to the classical \gls{WMMSE} algorithm at equivalent sum-rate performance in a dense downtown cell-site environment. Ultimately, this work bridges the gap between the strong channel-modeling capabilities of Transformer architectures and the stringent power budgets of next-generation 6G edge hardware.

\section*{Acknowledgement}
This work was supported by Ericsson - Global Artificial Intelligence Accelerator AI-Hub Canada in Montr\'{e}al and jointly funded by NSERC Alliance Grant 566589-21 (Ericsson, ECCC, Innov\'{E}\'{E}).

\bibliographystyle{IEEEtran}
\bibliography{0_main}

\end{document}